\documentclass[conference]{IEEEtran}
\IEEEoverridecommandlockouts

\usepackage{cite}
\usepackage{booktabs}
\usepackage{amsmath,amssymb,amsfonts}
\usepackage{algorithmic}
\usepackage{graphicx}
\usepackage{textcomp}
\usepackage{xcolor}
\usepackage{authblk}
\usepackage{hyperref}

\def\BibTeX{{\rm B\kern-.05em{\sc i\kern-.025em b}\kern-.08em
    T\kern-.1667em\lower.7ex\hbox{E}\kern-.125emX}}

\begin{document}

\title{A Full-System Simulation Framework for CXL-Based SSD Memory System\\

}

\author[1]{Yaohui Wang}
\author[1]{Zicong Wang}
\author[1]{Fanfeng Meng}
\author[1]{Yanjing Wang}
\author[1]{Yang Ou}
\author[1]{Lizhou Wu}
\author[1]{\\Wentao Hong}
\author[1]{Xuran Ge}
\author[1]{Jijun Cao}

\affil[1]{National University of Defense Technology,
Changsha, China \authorcr
  \{wangyaohui, wangzicong, mengfanfeng, wangyanjing, ouyang06, lizhou.wu,  \authorcr
  hongwt22, gexuran2, caojijun0\}@nudt.edu.cn}
\maketitle

\begin{abstract}
\textit{Compute eXpress Link (CXL)} is a promising technology for memory disaggregation and expansion. Especially, CXL makes it more effectively for large-capacity storage devices such as \textit{Solid State Drive (SSD)} to be deployed in the memory pool. However, CXL-based SSDs are still in early stages, necessitating the development of reliable simulation tools. In this paper, we propose \textbf{CXL-SSD-Sim}, the first open-source full-system simulator designed to simulate CXL-based SSD memory system. Constructed on the foundation of \textit{gem5} and \textit{SimpleSSD}, CXL-SSD-Sim extends an high fidelity SSD memory expander model along with the corresponding device driver. In addition, CXL-SSD-Sim models a DRAM layer as a caching mechanism for the SSD, meticulously engineered to counteract latency issues inherent to CXL-based SSD memory access. Experiments are performed among five different memory devices with CXL-SSD-Sim in aspect of latency, bandwidth and real-world benchmark performance. These experiments serve to underscore the efficacy of our simulation tool in providing a comprehensive analysis of CXL-based SSD memory systems. The CXL-SSD-Sim simulator is available at \href{https://github.com/WangYaohuii/CXL-SSD-Sim}{https://github.com/WangYaohuii/CXL-SSD-Sim}.
\end{abstract}


\section{Introduction}
With the growing popularity of massive data-driven applications such as large language model (LLM), the complexity and scale of deep learning models are increasing exponentially, leading to a surge in server memory demand\cite{memory_wall2,deeplearning}. The explosive growth in data volume, combined with the slowing of \textit{Moore's Law}, widens the gap between processor computational power and memory bandwidth, creating a \textit{memory wall} that hinders system performance\cite{memory_wall1, memory_wall2}. The storage requirements for model training and feature vectors far exceed the DRAM capacity of conventional server, with individual models occupying tens of terabytes of space. This shortage in memory capacity stifles the potential of data-intensive applications.
% RDMA-based pooled memory technologies, central to current data center infrastructures, offer high-performance caching and distributed storage solutions\cite{2024rcmp_wang, 2020effectively_al, 2019software_lagar}. However, their performance bottlenecks and cost-effectiveness are increasingly scrutinized. 
To address the server memory capacity bottleneck, many researches have explored using \textit{Solid State Drive (SSD)} as extended memory\cite{Hybrid, Flatflash, FlashVM, memory_wall1, SSD,pmem}.
% Companies like Baidu and NVIDIA have implemented SSD for storing preprocessed AI training results, utilizing a two-tier caching mechanism to reduce frequent accesses and prolong SSD lifespan while enhancing access efficiency. However, this adds complexity to software design and imposes high demands on data migration control to SSD.

Memory disaggregation technologies have emerged, designed to decouple memory resources from processing units, thereby enhancing their utilization and minimizing wastage \cite{efficient,cxl_expansion,memory_disaggregation}. The \textit{Compute Express Link (CXL)} \cite{cxl}, an emerging hardware interconnect protocol, showcases immense potential for enhancing memory pool system performance with its low-latency, high-bandwidth, and cache-coherent features \cite{cxl_mem}. CXL enables the connection of large-capacity storage devices like SSD to memory pools \cite{SSD}, offering a viable alternative to overcome DRAM's cost and capacity constraints. Its byte-level addressing capability permits direct CPU access to remote memory through \textit{load/store} instructions \cite{cxl}, minimizing traditional DMA-induced cache coherence overhead and drastically reducing memory access latency.

However, CXL is in its infancy stage, plagued by high prototype costs and a scarcity of mature devices \cite{direct_access,SSD,pond,gouk2024breaking,sun2023demystifying,tang2024exploring}. Existing CXL memory pool researches often relie on NUMA node simulation or specialized hardware, which suffers from insufficient simulation accuracy and high hardware customization difficulties. Thus, a comprehensive full-system simulation tool that accurately models the intricate interactions between processors, memory networks, and storage hardware is essential for analyzing and optimizing CXL memory pool systems.

Our proposed CXL-SSD-Sim, built on \textit{gem5} \cite{gem5} and \textit{SimpleSSD} \cite{SimpleSSD}, fills this gap, offering an open-source platform for researchers to test CXL memory devices performance and address the high-capacity memory needs of AI model training and key-value stores through CXL-compatible SSD storage. Yet, SSD integrated into CXL face high average access latency issues \cite{latency}. Despite exceptional read/write speeds nearing PCI Express 4.0 bus theoretical bandwidth, SSD exhibit latency in the microseconds to tens of microseconds range, significantly higher than DRAM's nanosecond response times. This delay characteristic directly affects the execution efficiency of CPU. To mitigate this, we optimized CXL-SSD-Sim by adding a DRAM cache layer, enhancing storage capacity while maintaining low latency.

To the best of our knowledge, CXL-SSD-Sim is the first open-source full-system CXL-based SSD memory simulator. The primary contributions
of this paper are listed below.
\begin{itemize}
    \item We built a full-system CXL-based SSD memory expander model along with the corresponding device driver for system-level simulation research.
    \item We developed an enhanced CXL-based SSD memory expander model which integrates a DRAM cache layer to reduce the memory access latency of SSD and improve their endurance. This provides flexibility to explore the architecture of CXL-based SSD memory system for researchers.
    \item A series of experimental evaluations have been conducted utilizing the CXL-SSD-Sim simulation platform to assess five different memory devices. The findings from these experiments are instrumental in substantiating the utility and accuracy of CXL-SSD-Sim as a research tool, particularly within the purview of current academic and developmental inquiries.
\end{itemize}    

\section{CXL-SSD-Sim Design and Implementation}
 On the basis of gem5 and SimpleSSD, we added CXL protocol parsing module and cache module to optimize the performance of memory system with SSD, and we have finished the CXL device driver and CXL protocol design work.

\subsection{CXL-SSD-Sim Overview Architecture}
\begin{figure}[ht]
  \centering
  \includegraphics[width=\linewidth]{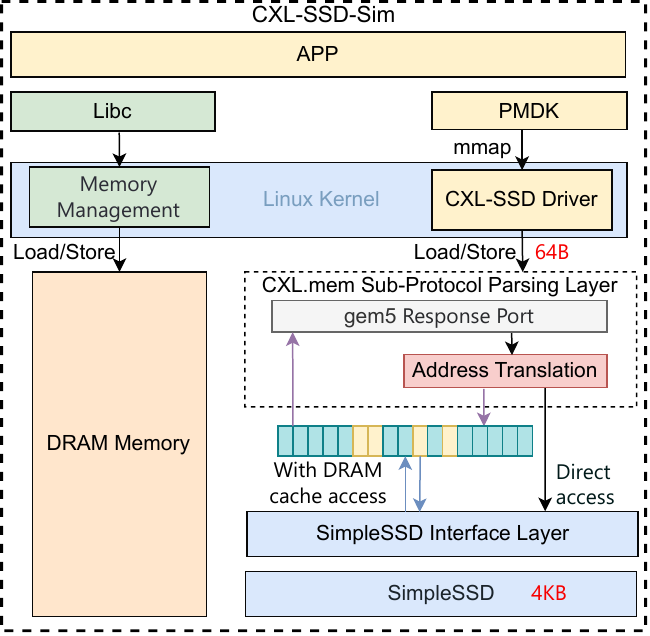}
  \caption{ Overall structure of the CXL-SSD-Sim extended memory device simulator.}
  \label{fig:overview}
\end{figure}

The basic structure of the CXL-SSD device is shown in the Fig.~\ref{fig:overview}. At the top level, applications (APP) interact with the Memory Management via \textit{Libc} and \textit{PMDK} (Persistent Memory Development Kit) libraries \cite{pmdk}. The Memory Management unit, located in the Linux kernel, manages memory resources. We designed a CXL-SSD kernel driver, mapping CXL devices to the Linux file system, allowing the CPU to access both local DRAM and the CXL-SSD-Sim simulator via load/store instructions. Additionally, we implemented a DRAM cache layer with five caching strategies for CXL-SSD-Sim.

The following components are central to our design:
\begin{itemize}
\item \textbf{CXL.mem sub-protocol parsing layer}: 
This layer includes packet format parsing, processing, CXL.mem sub-protocol parsing and request conversion to SimpleSSD access. CXL.mem sub-protocol parsing adds the processing of consistency information. In addition, it is necessary to extract the starting logical block address and the number of logical blocks from CXL Flit (64Byte), and combine them into the Request data structure in SimpleSSD.
\item \textbf{SimpleSSD interface layer}: This layer performs the following key functions: first, it invokes the SimpleSSD HIL::Read/Write interface to send requests to the SimpleSSD component. Second, the gem5 simulator determines the latency of access requests based on the Tick value returned by SimpleSSD. Additionally, the CXL-SSD model features an event-driven queue that interfaces with the abstract class provided by SimpleSSD to manage the device model. Upon gem5 initialization, the InitSimpleSSDEngine() function is called to set up the SimpleSSD CPU component and configure read operations.
\item \textbf{SimpleSSD simulator}:
SimpleSSD encompasses a complete storage stack, enabling a holistic performance evaluation that integrates various memory technologies and microarchitectures\cite{SimpleSSD}.
The interaction between the CXL-SSD device and SimpleSSD efficiently manages 64-byte cache line data but encounters challenges with 4KB logical blocks, leading to read-write amplification, which can increase storage traffic and impact latency.
\end{itemize}

\subsection{Implementation of CXL.mem Sub-protocol}
\subsubsection{CXL Memory Access}
\begin{figure}[ht]
  \centering
  \includegraphics[width=\linewidth]{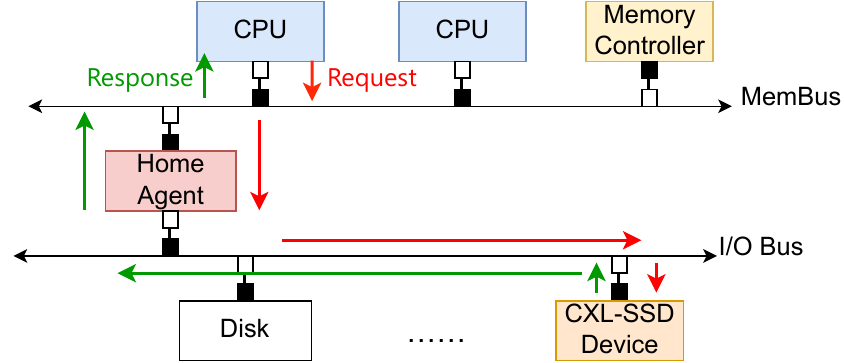}
  \caption{ CXL-SSD device access.}
  \label{fig:access_path}
\end{figure}

The memory access path for the CPU's load/store instructions is indicated by the red arrows in Fig.~\ref{fig:access_path}. The memory access response data path for CXL-SSD device is shown by the green arrows.
The memory access response data path for CXL-SSD device is shown by the green arrows.
Applications map virtual addresses to the CXL device using the OS's \textit{mmap} interface, and the CPU translates virtual to physical addresses via page tables.

When the CPU issues read/write instructions targeting the CXL-SSD memory device, these instructions are transformed into packets sent over the \textit{MemBus}. The \textit{Home Agent}, implemented as a \textit{Bridge} module in gem5, connects the system's MemBus and IOBus. Packets traversing the PCIe bus are converted into CXL.mem format, which encapsulates address ranges, cache line data, and coherence hints. The Home Agent handles packet address-port mapping, determining the physical address range targeted. If the packet is destined for CXL-SSD, it translates the packets into CXL Flits and forwards them to the device. The CXL-SSD then decodes the CXL.mem sub-protocol, processes the request, conducts packet format conversion, and sends a response back to the Home Agent.

\subsubsection{Packet Format Conversion and Address Mapping}
The Bridge intercepts packets targeting CXL-SSD memory, converting them to align with CXL standards. An extension in gem5's Packet class adds four CXL-specific transaction types:  M2S Request (M2SReq), M2S Request with Data(M2SRwD), S2M Data Response(S2MDRS), and S2M No Data Response(S2MNDR) \cite{cxl}. After conversion to a CXL Flit, latency associated with the protocol handling process is introduced in the Home Agent's event loop before forwarding the packet. For each packet passing through the Bridge, the Home Agent checks whether the target address belongs to the CXL extension memory device. If not, no packet format conversion occurs; otherwise, the request type is identified. Read requests (MemCmd::ReadReq) are converted into CXL.mem read transactions (MemCmd::M2SReq), and write requests (MemCmd::WriteReq) are converted into CXL.mem write transactions (MemCmd::M2SRwD). Other requests trigger a warning.

\subsubsection{Consistency Field Handling in CXL.mem}
The M2S Request message in CXL.mem includes a MetaValue field, which conveys consistency information about whether the host has a copy of the data at the specified address. \textit{Invalid} indicates that the host does not have a cacheable copy of the cache line. \textit{Any} indicates that the host may have a cache line copy in either shared, exclusive, or modified state. \textit{Shared} indicates that the host retains at least one cache line copy in a shared state.

In the Bridge, the conversion logic from gem5 Packet to CXL.mem sub-protocol handles the consistency field based on the Req field. If the Packet doesn't invalidate or flush the cache line, MetaValue is set to Any. If it invalidates, MetaValue is set to Invalid; if it flushes the cache without invalidating it, MetaValue is Shared.

\subsection{Cache Design of CXL-SSD-Sim}
To address the granularity mismatch in CXL-SSD and high SSD latency, a DRAM cache layer is used. This helps buffering SSD data and improving performance. The DRAM cache in CXL-SSD-Sim uses 4KB pages with dirty and valid bits, following a write-back, write-allocate policy. The cache replacement module manages page hits or evictions based on 4KB page numbers.

The DRAM cache mitigates the mismatch between 64B cache lines and 4KB SSD pages. The MSHR module handles overlapping 64B requests targeting the same 4KB page, avoiding redundant SSD reads and reducing data traffic. If the L2 cache hits the DRAM cache, CXL-SSD-Sim achieves latency close to CXL-based DRAM. In addition, CXL-SSD-Sim supports five caching strategies: Direct Mapping (Direct), Least Recently Used (LRU), First-In First-Out (FIFO), Two Queues (2Q), and Least Frequently Recently Used (LFRU).

\section{Experiments and Evaluation}
We evaluate various memory devices in CXL-SSD-Sim, focusing on latency, bandwidth, and throughput in persistent storage applications. Devices include DRAM, CXL-DRAM, persistent memory (PMEM), CXL-SSD without cache, and CXL-SSD with cache.

\subsection{Experimental Setup}
The experiment parameters are summarized in Table~\ref{tab:commands}. SimpleSSD is also officially verified. Therefore, the accuracy of the simulator can be confirmed. 
PMEM persistent memory device parameters follow \textit{SpecPMT} \cite{2023specpmt_ye}, with 150ns read and 500ns write latency. CXL.mem sub-protocol processing latency is 25ns (which is referenced by \cite{set_latency}). Total CXL.mem latency within the memory access network is 50ns (which is validated by real hardware testing with our FPGA-based CXL memory prototype \cite{cxl_ip}). DRAM cache access latency for CXL-SSD is 50ns. The primary system memory is 512MB DDR4 with one channel, and a single CPU core is used.

\begin{table}
\caption{CXL-SSD-Sim Experimental Environment Parameter Configuration}
\begin{center}
\label{tab:commands}
\scalebox{0.8}{
\begin{tabular}{clcl}
\hline
    \toprule
    \textbf{Parameter} & \textbf{Configuration} & \textbf{Parameter} & \textbf{Configuration}\\
\hline
    \midrule
    CPU type&x86&mem type&DDR4\_2400\_8x8   \\
    LQEntries/SQEntries&256&memory channels&1 \\
    numROBEntires&512&cpu number&1\\
    L1cache write\ buffers&128&PMEM\ rowbuffer&256B\\
    L1dcache size&64KB&PMEM read&150ns \\
    L1icache size& 32KB &PMEM write& 500ns \\
    CXL.mem latency&25ns &DRAM cache capacity& 16MB\\ 
    L2cache write\ buffers& 256 &SSD capacity& 16GB \\
    L2cache size& 512KB\\
    \bottomrule
\hline
\end{tabular}
}
\end{center}
\end{table}

\subsection{Latency and Bandwidth Experiment}
\begin{figure}[ht]
  \centering
  \includegraphics[width=\linewidth]{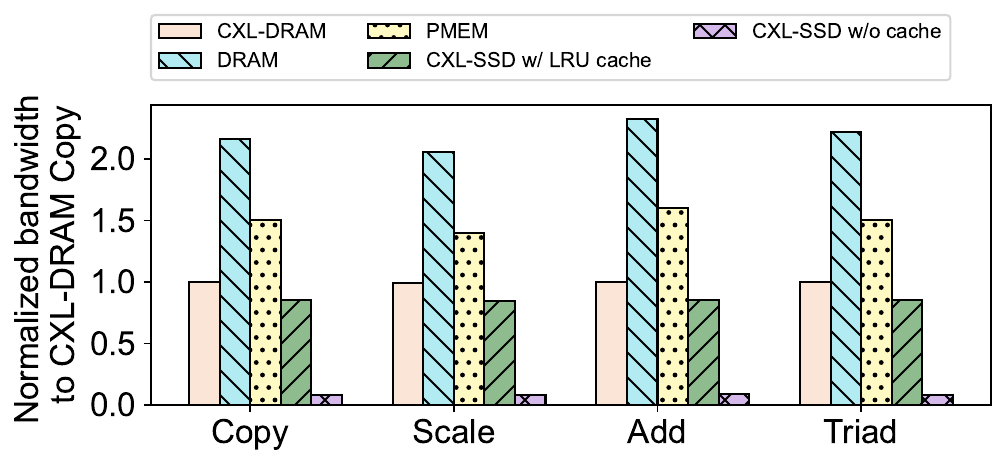}
  \caption{Bandwidth performance comparison across different memory devices.}
  \label{fig:bandwidth}
\end{figure}

\begin{figure}[ht]
  \centering
  \includegraphics[width=\linewidth]{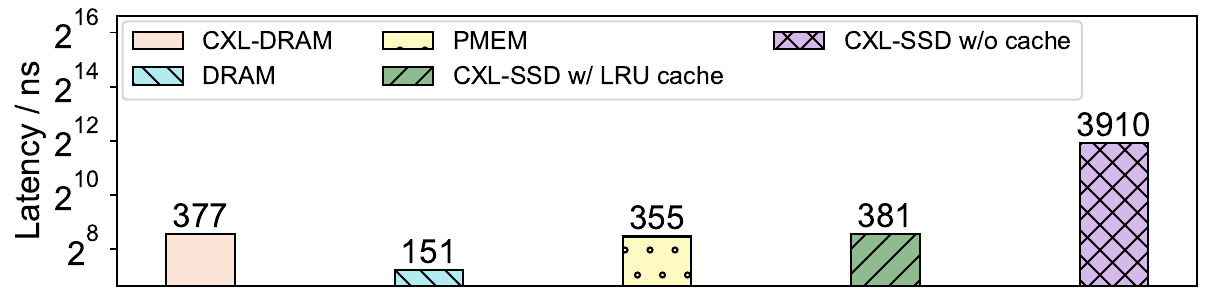}
  \caption{Latency performance comparison across different memory devices.}
  \label{fig:latency}
\end{figure}

We used \textit{stream} \cite{stream} to measure bandwidth and \textit{membench} \cite{membench} to assess latency on our simulator, comparing CXL-DRAM, DRAM, PMEM, CXL-SSD, and CXL-SSD with LRU cache, as shown in Fig.~\ref{fig:bandwidth} and Fig.~\ref{fig:latency}. The test employed an 8MB dataset for bandwidth and random read memory access for latency.

As shown in Fig.~\ref{fig:bandwidth}, DRAM demonstrated the highest bandwidth among the four operations tested, while CXL-SSD with an LRU cache achieved performance similar to CXL-DRAM. PMEM's bandwidth reached approximately 65\% of DRAM's.

Although CXL memory devices (CXL-DRAM and CXL-SSD) have higher latencies than DRAM due to the greater distance to the CPU, the CXL protocol allows them to expand CPU memory through unified addressing. When the DRAM cache is enabled, CXL-SSD exhibit performance on par with CXL-DRAM and PMEM. When the DRAM cache is not enabled, the average latency of SSD operations is excessively high, resulting in significantly longer delays compared to accessing hot data in DRAM with caching enabled. Thus, adding a cache layer to CXL-SSD can reduce traffic to the flash memory backend, improving overall performance.

\subsection{Application Experiment}

\begin{figure}[ht]
  \centering
  \includegraphics[width=\linewidth]{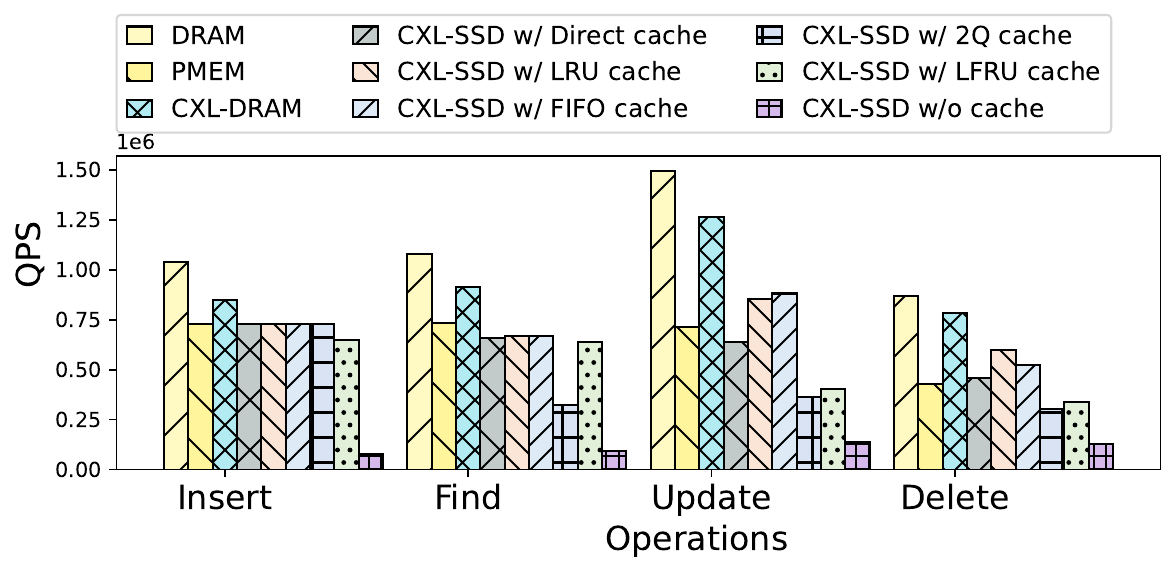}
  \caption{QPS comparison for Viper 216B key-value pairs across different memory devices.}
  \label{fig:viper216B}
\end{figure}

\textit{Viper}, a key-value store testing tool, was used to evaluate different memory devices, including CXL-SSD with various cache replacement policies \cite{viper}. Experiments involved key-value pairs of 216B and 532B sizes, performing 10,000 write, insert, query, update, and delete operations. Results for the 216B tests are shown in Fig.~\ref{fig:viper216B}.

DRAM and CXL-DRAM show notably higher throughput than PMEM and CXL-SSD, with CXL-DRAM experiencing only a 14\% performance loss compared to DDR4 DRAM. PMEM's performance falls 20-50\% behind CXL-DRAM, and CXL-SSD with LRU cache outperforms its non-cached version by 7 to 10 times on average.

\begin{figure}[ht]
  \centering
  \includegraphics[width=\linewidth]{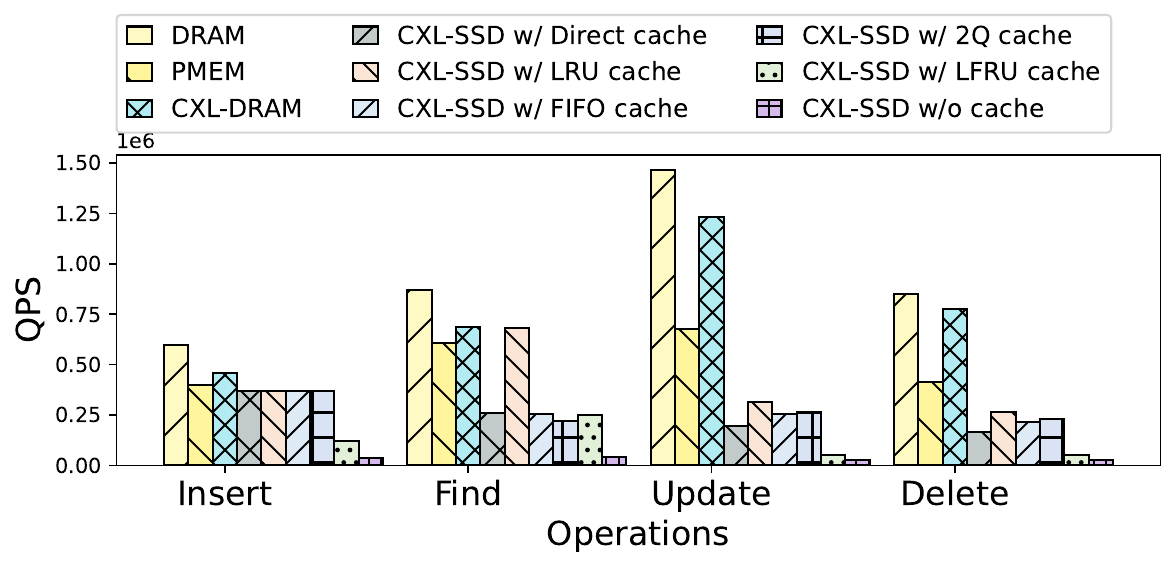}
  \caption{QPS comparison for Viper 532B key-value pairs across different memory devices.}
  \label{fig:viper532B}
\end{figure}

In Fig.~\ref{fig:viper532B}, for 532-byte key-value pairs, the performance (QPS) decreases as data size increases. DRAM and CXL-DRAM maintain higher performance, while PMEM's update and delete operations achieve 50\% of DRAM's speed. The CXL-SSD with cache suffers from higher cache miss rates as data granularity increases, leading to a 20-30\% degradation in QPS compared to PMEM.

The cache hit rate of various caching strategies in the \textit{viper} program is influenced by its memory access patterns. The program exhibits high temporal locality, particularly during update and delete operations, leading to repeated metadata access. Among the five caching strategies, LRU performs best. CXL-SSD throughput is strongly correlated with DRAM cache hit rate, while 2Q performs poorly. In high temporal locality scenarios, FIFO reduces LRU's effective cache space, causing performance degradation. A higher CXL-SSD cache hit rate compared to PMEM lowers average read-write latency, improving overall performance.
For different types of operations, write requests are generated by insert, update, and delete operations. If the hit rate of CXL-SSD is high, the write latency of the DRAM cache layer will be significantly lower than that of PMEM, making the performance of the CXL-SSD superior to that of the PMEM.

\section{Conclusion}
CXL is a promising technology for memory disaggregation, allowing efficient integration of large-capacity storage like SSDs into memory systems. 
Our work is inspired by Expander\cite{Kwon}. In order to promote the academic research of CXL, we introduce CXL-SSD-Sim, the first open-source full-system simulation framework built on gem5 and SimpleSSD, designed to model CXL-based SSD memory systems. It includes a high-fidelity SSD memory expander model the corresponding device driver. Furthermore, we introduce a DRAM layer, designed as a caching mechanism for the SSD, which is meticulously optimized to mitigate latency challenges intrinsic to CXL-based SSD memory access, thereby enhancing access efficiency and extending the operational lifespan of the SSD. We conduct experiments using five different memory devices, evaluating latency, bandwidth, and real-world benchmarks via CXL-SSD-Sim. Compared with the trace-based simulator MQSim \cite{overcoming}, our simulator has comprehensive system simulation capabilities. Using the features of gem5 full system, it can not only run simple benchmarks or specific applications, but also support modifying and testing the operating system, and analyze the real software stack.

\bibliographystyle{IEEEtran}
\bibliography{conference_101719}

\begin{thebibliography}{10}
\providecommand{\url}[1]{#1}
\csname url@samestyle\endcsname
\providecommand{\newblock}{\relax}
\providecommand{\bibinfo}[2]{#2}
\providecommand{\BIBentrySTDinterwordspacing}{\spaceskip=0pt\relax}
\providecommand{\BIBentryALTinterwordstretchfactor}{4}
\providecommand{\BIBentryALTinterwordspacing}{\spaceskip=\fontdimen2\font plus
\BIBentryALTinterwordstretchfactor\fontdimen3\font minus \fontdimen4\font\relax}
\providecommand{\BIBforeignlanguage}[2]{{%
\expandafter\ifx\csname l@#1\endcsname\relax
\typeout{** WARNING: IEEEtran.bst: No hyphenation pattern has been}%
\typeout{** loaded for the language `#1'. Using the pattern for}%
\typeout{** the default language instead.}%
\else
\language=\csname l@#1\endcsname
\fi
#2}}
\providecommand{\BIBdecl}{\relax}
\BIBdecl

\bibitem{memory_wall2}
G.~Amir, Y.~Zhewei, S.~Kim, W.~M. Michael, and K.~Keutzer, ``Ai and memory wall,'' \emph{RiseLab Medium Post}, 2021.

\bibitem{deeplearning}
D.~Mudigere, Y.~Hao \emph{et~al.}, ``Software-hardware co-design for fast and scalable training of deep learning recommendation models,'' in \emph{ISCA '22: Proceedings of the 49th Annual International Symposium on Computer Architecture}, 2022.

\bibitem{memory_wall1}
J.~Huang, A.~Badam, M.~K. Qureshi, and K.~Schwan, ``Unified address translation for memory-mapped ssds with flashmap,'' in \emph{Proceedings of the 42Nd Annual International Symposium on Computer Architecture}, 2015, pp. 580--591.

\bibitem{Hybrid}
A.~Badam and V.~S. Pai, ``Ssdalloc: Hybrid ssd/ram memory management made easy,'' in \emph{8th USENIX Symposium on Networked Systems Design and Implementation (NSDI 11)}, 2011.

\bibitem{Flatflash}
A.~Abulila, V.~S. Mailthody, Z.~Qureshi, J.~Huang, N.~S. Kim, J.~Xiong, and W.-m. Hwu, ``Flatflash: Exploiting the byte-accessibility of ssds within a unified memory-storage hierarchy,'' in \emph{Proceedings of the Twenty-Fourth International Conference on Architectural Support for Programming Languages and Operating Systems}, 2019, pp. 971--985.

\bibitem{FlashVM}
M.~Saxena and M.~M. Swift, ``Flashvm: Virtual memory management on flash,'' in \emph{2010 USENIX Annual Technical Conference (USENIX ATC 10)}, 2010.

\bibitem{SSD}
M.~Jung, ``Hello bytes, bye blocks: Pcie storage meets compute express link for memory expansion (cxl-ssd),'' in \emph{Proceedings of the 14th ACM Workshop on Hot Topics in Storage and File Systems}, 2022, pp. 45--51.

\bibitem{pmem}
J.~Yang, J.~Kim, M.~Hoseinzadeh, J.~Izraelevitz, and S.~Swanson, ``An empirical guide to the behavior and use of scalable persistent memory,'' in \emph{18th USENIX Conference on File and Storage Technologies (FAST 20)}, 2020, pp. 169--182.

\bibitem{efficient}
J.~Gu, Y.~Lee, and Y.~Zhang, ``Efficient memory disaggregation with infiniswap,'' in \emph{14th USENIX Symposium on Networked Systems Design and Implementation (NSDI 2017)}, 2017.

\bibitem{cxl_expansion}
M.~Ahn, A.~Chang, D.~Lee, J.~Gim, J.~Kim, J.~Jung, O.~Rebholz, V.~Pham, K.~Malladi, and Y.~Ki, ``Enabling cxl memory expansion for in-memory database management systems,'' in \emph{DaMoN'22: Data Management on New Hardware}, 2022.

\bibitem{memory_disaggregation}
H.~A. Maruf and M.~Chowdhury, ``Memory disaggregation: Advances and open challenges,'' \emph{ACM SIGOPS Operating Systems Review}, pp. 29--37, 2023.

\bibitem{cxl}
\BIBentryALTinterwordspacing
Cxl consortium. [Online]. Available: \url{https:// www.computeexpresslink.org/}
\BIBentrySTDinterwordspacing

\bibitem{cxl_mem}
Y.~Fridman, S.~Mutalik~Desai, N.~Singh, T.~Willhalm, and G.~Oren, ``Cxl memory as persistent memory for disaggregated hpc: A practical approach,'' in \emph{SC-W 2023: Workshops of The International Conference on High Performance Computing, Network, Storage, and Analysis}, 2023.

\bibitem{direct_access}
D.~Gouk, S.~Lee, M.~Kwon, and M.~Jung, ``Direct access,high-performance memory disaggregation with directcxl,'' in \emph{2022 USENIX Annual Technical Conference (USENIX ATC 22)}, 2022, pp. 287--294.

\bibitem{pond}
\BIBentryALTinterwordspacing
H.~Li, D.~S. Berger, L.~Hsu, D.~Ernst, P.~Zardoshti, S.~Novakovic, M.~Shah, S.~Rajadnya, S.~Lee, I.~Agarwal, M.~D. Hill, M.~Fontoura, and R.~Bianchini, ``Pond: Cxl-based memory pooling systems for cloud platforms,'' in \emph{Proceedings of the 28th ACM International Conference on Architectural Support for Programming Languages and Operating Systems, Volume 2}, ser. ASPLOS 2023.\hskip 1em plus 0.5em minus 0.4em\relax New York, NY, USA: Association for Computing Machinery, 2023, p. 574–587. [Online]. Available: \url{https://doi.org/10.1145/3575693.3578835}
\BIBentrySTDinterwordspacing

\bibitem{gouk2024breaking}
D.~Gouk, S.~Kang, H.~Bae, E.~Ryu, S.~Lee, D.~Kim, J.~Jang, and M.~Jung, ``Breaking barriers: Expanding gpu memory with sub-two digit nanosecond latency cxl controller,'' in \emph{Proceedings of the 16th ACM Workshop on Hot Topics in Storage and File Systems}, 2024, pp. 108--115.

\bibitem{sun2023demystifying}
Y.~Sun, Y.~Yuan, Z.~Yu, R.~Kuper, C.~Song, J.~Huang, H.~Ji, S.~Agarwal, J.~Lou, I.~Jeong \emph{et~al.}, ``Demystifying cxl memory with genuine cxl-ready systems and devices,'' in \emph{Proceedings of the 56th Annual IEEE/ACM International Symposium on Microarchitecture}, 2023, pp. 105--121.

\bibitem{tang2024exploring}
Y.~Tang, P.~Zhou, W.~Zhang, H.~Hu, Q.~Yang, H.~Xiang, T.~Liu, J.~Shan, R.~Huang, C.~Zhao \emph{et~al.}, ``Exploring performance and cost optimization with asic-based cxl memory,'' in \emph{Proceedings of the Nineteenth European Conference on Computer Systems}, 2024, pp. 818--833.

\bibitem{gem5}
\BIBentryALTinterwordspacing
gem5. [Online]. Available: \url{https://www.gem5.org/}
\BIBentrySTDinterwordspacing

\bibitem{SimpleSSD}
\BIBentryALTinterwordspacing
Simplessd. [Online]. Available: \url{https://docs.simplessd.org/en/v2.0.12/}
\BIBentrySTDinterwordspacing

\bibitem{latency}
R.~H. Arpaci-Dusseau and A.~C. Arpaci-Dusseau, ``Operating systems: Three easy pieces,'' 2018.

\bibitem{pmdk}
\BIBentryALTinterwordspacing
Pmdk. [Online]. Available: \url{https://www.intel.cn/content/www/cn/zh/developer/topic-technology/persistent-memory/overview.html}
\BIBentrySTDinterwordspacing

\bibitem{2023specpmt_ye}
C.~Ye, Y.~Xu, X.~Shen, Y.~Sha, X.~Liao, H.~Jin, and Y.~Solihin, ``Specpmt: Speculative logging for resolving crash consistency overhead of persistent memory,'' in \emph{Proceedings of the 28th ACM International Conference on Architectural Support for Programming Languages and Operating Systems, Volume 2}, 2023, pp. 762--777.

\bibitem{set_latency}
D.~D. Sharma, ``Compute express link{\textregistered}: An open industry-standard interconnect enabling heterogeneous data-centric computing,'' in \emph{2022 IEEE Symposium on High-Performance Interconnects (HOTI)}.\hskip 1em plus 0.5em minus 0.4em\relax IEEE, 2022, pp. 5--12.

\bibitem{cxl_ip}
\BIBentryALTinterwordspacing
Intel cxl ip. [Online]. Available: \url{https://www.intel.com/content/www/us/en/ products/details/fpga/intellectual-property/interface-protocols/cxl-ip.html}
\BIBentrySTDinterwordspacing

\bibitem{stream}
\BIBentryALTinterwordspacing
Stream: Sustainable memory bandwidth in high performance computers. [Online]. Available: \url{https://www.cs.virginia.edu/stream}
\BIBentrySTDinterwordspacing

\bibitem{membench}
N.-A. Tehrany, ``Evaluating performance characteristics of the pmdk persistent memory software stack,'' Ph.D. dissertation, Vrije Universiteit Amsterdam, 2020.

\bibitem{viper}
L.~Benson, H.~Makait, and T.~Rabl, ``Viper: An efficient hybrid pmem-dram key-value store,'' 2021.

\bibitem{Kwon}
M.~Kwon, S.~Lee, and M.~Jung, ``Cache in hand: Expander-driven cxl prefetcher for next generation cxl-ssds,'' in \emph{15th ACM Workshop on Hot Topics in Storage and File Systems (HotStorage)}, 2023.

\bibitem{overcoming}
S.-P. Yang, M.~Kim, S.~Nam, J.~Park, J.-Y. Choi, E.~H. Nam, E.~Lee, S.~Lee, and B.~S. Kim, ``Overcoming the memory wall with cxl-enabled ssds,'' in \emph{2023 USENIX Annual Technical Conference (USENIX ATC 23)}, 2023, pp. 601--617.

\end{thebibliography}

\vspace{12pt}
\end{document}